\UseRawInputEncoding
\documentclass[preprint,dvips,numbers,sort&compress]{elsarticle}
\usepackage{amsmath}
\usepackage{amssymb}
\usepackage{amsfonts}
\usepackage{amsthm}
\usepackage{arydshln}
\usepackage[toc, page]{appendix}
\usepackage{vector}
\usepackage{graphicx}
\usepackage{subfigure}
\usepackage{color}
\usepackage{paralist}
\usepackage{multirow}
\usepackage{booktabs}
\usepackage{setspace}
\graphicspath{{xia_eps/}}

\numberwithin{equation}{section}
\begin{document}
\title{ 
Digest Ultrasonic Welding I: Localized Heating and Fuse Bonding}
\author[a1]{Lixiang Yang}
\address[a1]{Department of Mechanical and Aerospace Engineering, The Ohio State University, Columbus OH, 43210 Email: yang.1130@buckeyemail.osu.edu}
\author[a2]{Libin Yang}
\address[a2]{ 
School of Industrial Automation, Beijing Institute of Technology, Zhuhai, 519088, China
Email: 12261@bitzh.edu.cn}
%

\begin{abstract} \noindent
This paper reports a new insight on the principle of ultrasonic welding process.  Physics of ultrasonic welding process is considered to be an inverse process of shear band formation. Both ultrasonic welding and shear band formation involve temperature rising due to localized plastic deformation. During shear band formation, heat will lead to damage of materials, e.g., initiation of cracks inside materials. But heat generated in ultrasonic welding process will bond two pieces or repair the cracks inside materials \citep{barroeta_robles_repair_2022}. Energy directors used in ultrasonic welding process act as a trigger of localized plastic deformation. Mathematically, We found out that heating generation during ultrasonic welding process is similar to heating generation by electric current.

\end{abstract}
\maketitle
\section{Introduction}
\label{s:intro} \noindent
Materials with high strength and high toughness are the idea materials for many industry applications. People are consistently seeking high strength and high toughness materials \citep{sun_highly_2012, gong_double-network_2003}. High strength implies high Young's modulus and yielding stress. High toughness means materials can absorb a lot of mechanical energy before fracture. Fracture of materials corresponds to crack formation in a piece of structure, that is, one piece becomes two pieces. Before occurs of material fracture, a plastic cohesive zone will usually form \citep{ahmadi_experimental_2022}. When plastic deformation  become localized, shear bands or neckings will be formed. These shear bands and neckings will be generated at the weakest point in the materials, accompanying with localized heating. This generated thermal energy will act as a destroying force to molecular bonds. When thermal energy is greater than bonding energy of molecular, fracture will be possible \citep{slootman_quantifying_2020}. Part of internal energy dissipates to become surface energy\citep{zhou_high-throughput_2022}. 

On the other hand, ultrasonic welding is widely used to bond similar and dissimilar polymers, metals, and woods\citep{unnikrishnan_review_2022,li_ultrasonic_2022, bendikiene_study_2021, jongbloed_differences_2020, wagner_ultrasonic_2013}.  When an ultrasonic welding device composed of converter, booster and sonotrode is used to bond two pieces of materials such as thermoplastics, thermoplastic elastomer, and thermosets, mechanical energy generated by ultrasound wave will be propagated, amplified and finally localized to the interface between two pieces of adherends \citep{yang_numerical_2022, rubino_ultrasonic_2020, guo_joining_2019}. Localization of stress wave energy will lead to thermal energy accumulated around the interface of adherends. Due to this thermal energy, the temperature rising rate will depend on thermal capacity of polymer materials since thermal capacities of polymers have a big shift across glass transition temperature \citep{koutras_thermal_2021}. Therefore, a large difference of temperature rising rate was observed across glass transition temperature\citep{zhang_study_2010}. After polymers absorb a large amount of thermal energy, molecular bonds will be broken and materials become flowable liquid shape. Once temperature is cooled down, molecular bonds will be reformed.

In summary, if a high speed impact force is applied on materials, one piece can be broken into two. If a cyclic vibration force is applied on materials,  two pieces can be stitched. Both are due to localized plastic deformation and accompanying temperature rising.

Traditionally, heating generation due to ultrasonic welding is attributed to frictional heating followed by viscoelastic heating \citep{zhang_study_2010, bhudolia_advances_2020,levy_modeling_2014}. That is on the right track. But it didn't tell the whole story of the ultrasonic welding process. Firstly, localized heating can hardly be explained by frictional heating and viscoelastic heating \citep{koellhoffer_role_2011, villegas_ultrasonic_2019}. Because frictional heating and viscoelastic heating happen everywhere inside materials, why only materials near the interface get heating up. If a longitudinal vibration is applied to adherends, frictional force on the interface can hardly be calculated. Secondly, with high temperature rising and large plastic deformation during ultrasonic vibration,  linear viscoelastic heating plays a minor role on bonding formation. Frictional heating is more related to molecular defect motions.  Recall in ultrasonic welding processes, energy directors are usually used to guarantee a good bonding quality. Physically, an energy director is used to direct mechanical energy to a particular location. It is directly related to how localized plastic energy changes to thermal energy. So energy directors act as mechanical impedances to block ultrasonic energy propagation. 

In the following section, we will demonstrate this process from physical aspects and mathematical aspects.

The rest of the present paper is organized as follows.  Section \ref{s:consti}
illustrates the physics of ultrasonic welding principle.  Section \ref{s:math}  is to explain a mathematical model of ultrasonic welding, especially temperature rising. 
Then we will talk about localized heating by using energy directors. Finally,  we offer the concluding remarks, followed by a list of cited references.   

\section{Physics of Ultrasonic Welding Principle} 
\label{s:consti} 
\noindent
It is well known that no materials are perfect. There are always some defects inside the materials. In polymers,  molecular chains are either cross-linked or entangled \citep{yang_note_2018}. de Gennes \citep{de_gennes_reptation_1971} gave an idea that reptation of molecular chains is caused by defect movements along a molecular chain. Equivalently, defects in molecular chains can be viewed as the results induced by frictional forces or entangled forces between different molecular chains. Plastic deformation in polymers is due to the accumulation and movement of these defects \citep{yang_mathematical_2019}. In metals, line defects such as dislocations are abundant. Dislocation densities for many metals are ranged from $10^{10}$ to $10^{16}/cm^2$. Dislocation generation and movement is closely related to plastic deformation. Therefore, for both polymers or metals, their defects will be the main causes for plastic deformation, especially localized plastic deformation and temperature rising. 

Nowadays, there are a lot of experimental investigations and numerical simulations of ultrasonic welding processes \citep{gohel_effect_2022, charlier_ultrasonic_2022, b_g_brito_effects_2022, frederick_disassembly_2021}. But work on understanding the principle of ultrasonic welding is relatively rare \citep{siddiq_thermomechanical_2008, li_integrated_2020}. 
In order to understand the principle of ultrasonic welding, firstly we need to understand energy flow due to ultrasonic wave. Particularly, we need to know the cases when stress wave can propagate \citep{yang_numerical_2010, yu_numerical_2010} and when stress wave will stop propagating. 
In general, when stress wave propagates, material particles will vibrate in their equilibrium positions. Mechanical energy will be transported to the next adjacent particle by molecular bonding forces. Wave propagation speed will depend on the stiffness of molecular bonding. Suppose there exists an ideal material which has no defects. Then energy will continue transporting to next particle without any loss. However, defects are always exist. Defects will act as impedances to energy transport.  For example, in metals, once energy carried by ultrasonic wave is absorbed by dislocations, immobile dislocations will be disengaged by this mechanical energy. Therefore, mobile dislocation density and mobile dislocation velocity will increase. The kinetic energy level of these mobile dislocations will increase as well. But they cannot stay in the higher energy state for a long time since they will collide with lattice frames. dislocation scattering will happen. Their kinetic energies will fall down to a lower energy level quickly.  After mobile dislocations lose their kinetic energy, they become immobile dislocations again. But immobile dislocation density will change in this energy gain and lose procedure. Accumulated immobile dislocation density will cause permanent plastic deformation. At the same time, because of mobile dislocation collision with lattice frames, dislocations lose kinetic energy. But lattice frames will gain kinetic energy. Lattices will vibrate more dramatically. Recall that temperature is a microscopic measurement of gas molecular velocity in idea gas or lattice vibration amplitude in solids. Therefore,  most lost kinetic energy of dislocations will change to lattice vibration which is macroscopically shown as a gained thermal energy. This leads to temperature rising.  Some of their energy will also radiate out as a sound wave.

Similarly, for polymers, because molecular chains are all entangled or cross-linked, bonding force of cross-links is much larger than entangled force. When energy carried by ultrasonic wave propagates along the molecular chains, entangled molecular chains can be disengaged by the ultrasonic wave energy e.g. many short molecular chains or danglings     
  \citep{lin_fracture_2021} can be formed. Therefore, mobile short molecular chain density and velocity will increase. Like dislocations, these short molecular chains will collide against main backbone chains. Collision scattering of these short molecular chains can cause them to lose their kinetic energy very quickly.  On the other hand, main backbone chains will gain kinetic energy from those discrete short molecular chains and vibrate more violently. This gives rise to temperature rising in the entire areas.
  Since energy from ultrasonic wave is firstly absorbed by these short chain defects, they will jump to a high energy level and return to a low energy level in a short time. 
 With redistribution of short chain density and velocity, plastic deformation, temperature rising and bonding will happen.

To make this physics more clear, mechanical energy flow could be compared with electric energy flow.
Physics of ultrasonic wave propagation in metals or polymers is similar to that of electrical current propagation in conductors. In electric circuits, when voltage is setup between a conductor, electrical current will carry electric energy from positive voltage to negative voltage. Due to existence of electric resistance, this electric energy transport is not lossless. Once voltage sets up an electric field inside conductor, electrons will gain a higher speed. Mobile electron density and velocity will increase. That is, these electrons will jump to a higher energy level. But these electrons can only stay at the higher energy level in a very short time, less than a microsecond. Because electron scatterings happen, high speed electrons will collide to nuclei and slow down. Kinetic energy of electrons will be lost. Some of their energy will dissipate as thermal energy and cause temperature rising. Some other will emit as  a light wave.

Furthermore, heat-induced shear band and heat-induced ultrasonic welding in materials can be analogy to electric fuse wire and electric soldering. 
For electric fuse wires, their melting temperature is much lower than conductor wires. Therefore, an electric fuse wire is the weakest point of the entire conductor wire when an electric current wave passes through. When a large amplitude electric current passes through an electric fuse, heat generated by a large electric current will melt the fuse wire and break it into two pieces. Similarly, if a large impact force is applied to metals or polymers, a high stress gradient and a large strain rate will be generated in materials. Mechanical energy carried by stress wave will pass through the materials. In the weakest point or the largest mechanical resistance area, mechanical energy will change to localized plastic and thermal energy. This leads to shear band and breaks a structure into two pieces.
In electric soldering process, electric energy is used to melt a solder. The solder will secure the connection of two metal surfaces. In ultrasonic welding process, mechanical energy is used to melt energy directors which will bond two pieces of materials.

\section{Mathematics of Ultrasonic Welding Principles} 
\label{s:math} 
\noindent

As mentioned before, ultrasonic welding is inverse to shear band and necking formation. Heat generated by impact force can damage materials and create cracks. But heat generated by ultrasonic wave can bond materials and eliminate cracks. Therefore, The same mathematical equation used to model shear band and necking formation can be applied to ultrasonic welding. 

\begin{figure}
\centering \includegraphics[scale = 0.3]{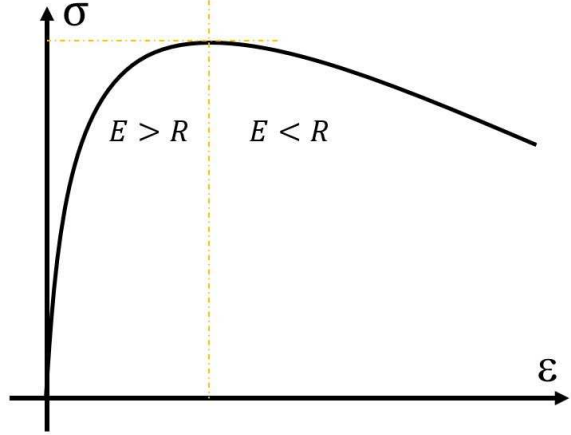}
\caption{
Sketch of a stress-strain relationship. Maximum point on the curve corresponds initiation of localized plastic deformation.
}

\label{f:f4}
\end{figure}

Yang and Yang \citep{yang_revisit_2020} proposed 
a mathematical model to explain localized plastic deformation such as shear band and necking with defects included in the equation. The mathematical structure can be written as

\begin{equation} 
\frac{d \sigma}{d \epsilon} \: = \: E \, - \, R \,  
\label{eq:localized} 
\end{equation}
where $ \sigma$ is engineering stress, $\epsilon$ is engineering strain, $E$ is Young's modulus, and $R$ is mechanical resistance. $R$ is a function of strain and stress and can be written as \citep{yang_revisit_2020}
\begin{equation} 
R\: = q\epsilon^m \sigma^n,  
\label{eq:resistance1} 
\end{equation}
where $E$ and $q$ are time, temperature, pressure dependent which can be modeled using modified logistic functions \citep{yang_viscoelasticity_2021}. $m$ and $n$ are constants depending on materials.
Eq.(\ref{eq:localized}) can be graphically illustrated as shown in Fig.(\ref{f:f4}).

Mechanical resistance $R$ increases with stress and strain, see Eq.(\ref{eq:resistance1}) and Young's modulus is independent of stress and strain. Therefore, the slope of stress and strain curve will gradually decrease with increased stress and strain. As shown in Fig.(\ref{f:f4}), the slope of stress and strain curve will be determined by the value of $E - R$ which can be positive or negative. Before maximum engineering stress is reached, e.g.,  $E > R$, slope is positive. Homogeneous elastic-plastic deformation will happen. After passing maximum engineering stress, e.g., $E < R$, slope becomes negative. Localized plastic deformation will happen. Localized thermal energy will accumulate in a very short band. Temperature in this narrow band can be higher than melting temperature of materials. 
Next, 
thermal energy due to ultrasonic wave will be illustrated.

Traditionally, energy transmission in ultrasonic welding is based on elastic wave propagation theory. In the interface between the sonotrode and the adherend, some of energy will transmit and some will reflect back \citep{palardy_study_2018}. Energy calculation is based on power reflection coefficient and transmission coefficient \citep{li_weld_2019, yang_modeling_2011}. With this method, mechanical energy arrived at the surface of adherend can be obtained \citep{rogale_interdependence_2022}. Next, this mechanical energy will enter the materials and change to thermal energy. How this mechanical energy changes to thermal energy is what we will discuss below. Currently most researchers explained this transition from mechanical energy to thermal energy by using linear viscoelastic theory \citep{yang_viscoelasticity_2013, yang_theoretical_2020}. Loss modulus is used to calculate the thermal energy. This approach may be acceptable in some cases such as small material deformation. The limitation of viscoelastic heating in traditional linear viscoelastic theory is that no micro-structures and defects of materials are considered in the model. In ultrasonic welding, most of ultrasonic wave energy will be absorbed by defects such as dislocations in metals or short molecular chains in polymers. These defect movements are related to nonlinear visco-plasticity. Large deformation, permanent deformation, large temperature rising can hardly be explained by linear viscoelasticity \citep{long_experiments_2022}. If only linear viscoelasticity is used to calculate temperature rising, temperature will not be high enough to melt the materials \citep{bakavos_mechanisms_2010}. For example, if ultrasonic frequency is 20000 Hz and bonding time is 1 ms, there are only 20 loading and unloading cycles. Nevertheless, localized temperature rising cannot be explained by linear viscoelasticity. Therefore, a new way to calculate temperature rising in ultrasonic welding is introduced. 
  
\begin{figure}
\includegraphics[scale = 0.2]{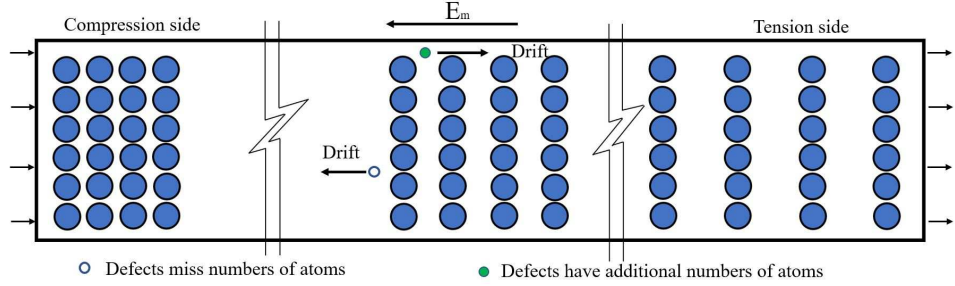}
\caption{
Schematic illustration of propagation of defects in crystalline materials. In compression side, atoms are closely packed. In tensile side, the spaces between atoms are wider. Mechanical potential is built based on   this uneven distribution of atoms. Defects such as dislocations will drift along this built-in potential
}

\label{f:f1}
\end{figure}

\begin{figure}
\includegraphics[scale = 0.2]{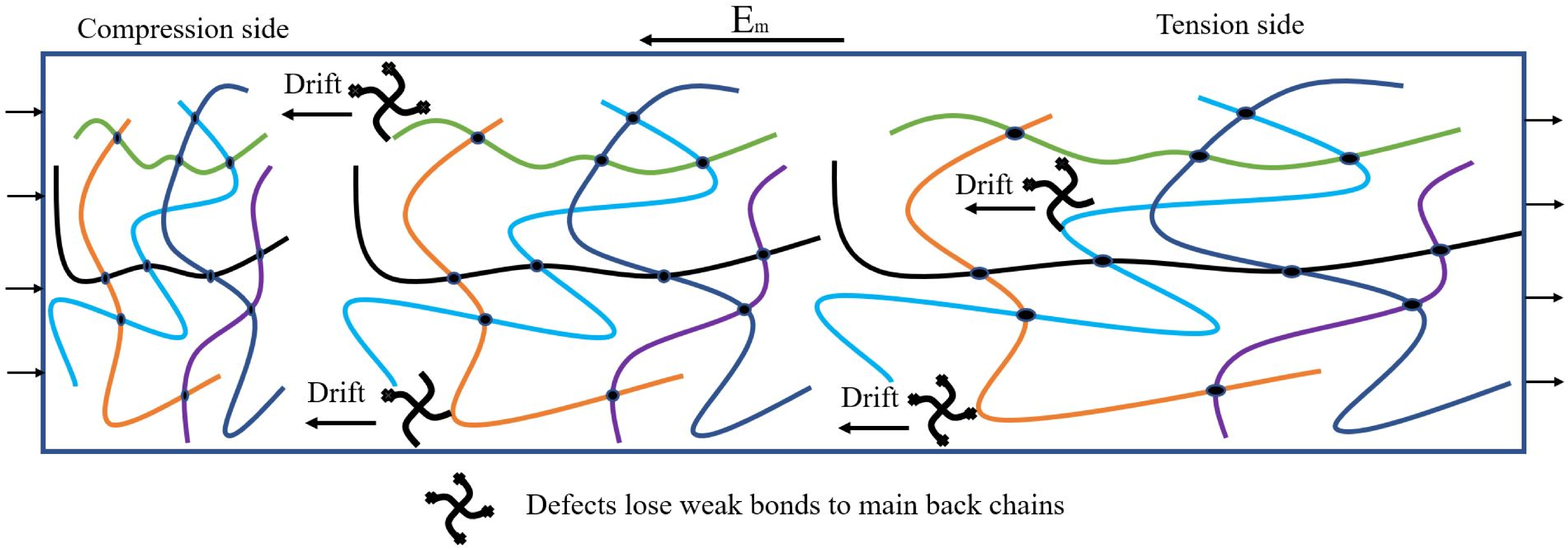}
\caption{
Schematic illustration of propagation of defects in polymer materials. In compression side, polymer chain density is high. In tensile side, polymer chain density is low. Mechanical potential is built based on   this uneven distribution of molecular chains. Defects such as danglings will drift along this built-in potential.
}
\label{f:f2}
\end{figure}

As we know, when electric potential is setup on a electric circuit, an electric field will be created. Electrons will drift in the electric field and create electric current $I$. Electric current is related to  charge density and charge velocity by
\begin{equation} 
I \: =\: nq_ev_eA,  
\label{eq:current} 
\end{equation}
where $n$ is electron concentration, $q_e=1.6 \times 10^{-19}$ coulombs/charge, $n \times q_e$ is electron density, $v_e$ is electron velocity, $A$ is cross section of conductor wires. The heat, $H$, generated by electric current in the electric resistance can be calculated by using Joule's law
\begin{equation} 
H \: =\: ( I ) \,^{2} R_e t,  
\label{eq:ohm2} 
\end{equation}
where $R_e$ is electric resistance. 
For dislocation induced plastic deformation, plastic strain rate, $\dot{\epsilon}_p$, is related to dislocation density and dislocation velocity by Orowan equation \citep{yang_mathematical_2019}  as
\begin{equation} 
\dot{\epsilon}_p \: = \rho b v,  
\label{eq:strainrate} 
\end{equation}
where $\rho$ is dislocation density, $b$ is Burger's vector, $v$ is dislocation velocity. 
By comparing with electric current equation, e.g., Eq.(\ref{eq:current}), we can define mechanical current due to dislocation drifting as
\begin{equation}
I_m = \dot{\epsilon}_p A_m = \rho b v A_m ,
\label{eq:mechanical_current}
\end{equation}
where $I_m$ is mechanical current and $A_m$ is the cross section of a metal bar.
If an electric potential $U$ is applied between an electric resistance $R_e$, a constant electric current $I$ will be maintained. Ohm's law is written as
\begin{equation} 
U \: = R_e I.  
\label{eq:ohm} 
\end{equation}
If a rod is under dynamical impact, stress inside the bar is not uniform. Consider a case that there is a stress gradient along the bar, e.g. one side is in tension while the other side is in compression, see Fig.\ref{f:f1}. There are more materials or equivalently atoms accumulated in the compression side than in the tension side which will build up a mechanical potential. This mechanical potential will setup a mechanical field. This mechanical field will drive dislocations to move along the field gradient direction. Dislocations are line defects which can be viewed as a combination of point defects. Point defects can be vacancies and interstitials. Dislocations with vacancies will drift to the compression side. Dislocations with interstitials will drift to the tension side. Similarly, for polymer bars, chain density in tension side is smaller than that in compression side. short chains who lose their connections to main chains will drift in the built-up field, see Fig.\ref{f:f2} for schematic illustrations.
To cast, if a mechanical potential $U_m$ is maintained between a rod with a mechanical resistance $R_m$, a constant mechanical current $I_m$ will be maintained.  Its relationship is written as
\begin{equation} 
U_m \: = R_m I_m.  
\label{eq:myohm} 
\end{equation}
Heat generated per unit volume can be calculated by
\begin{equation} 
H \: =\: ( \,  I_m ) \,^{2} R_m t,  
\label{eq:ohm1} 
\end{equation}
Analogy to electric resistance calculation method, mechanical resistance of materials can be obtained by
\begin{equation} 
R_m \: =\: l_m/\sigma_m A_m ,  
\label{eq:mech_resistance} 
\end{equation}
where $l_m$ is the length of mechanical resistor. $\sigma_m$ and $A_m$ are conductivity  and cross section of the mechanical resistor, respectively. Conductivity $\sigma_m$ will depend on dislocation scattering in metals or short molecular chain scattering in polymers.

\section{Energy Director}
\label{s:Energy}
\begin{figure}
\includegraphics[scale = 0.2]{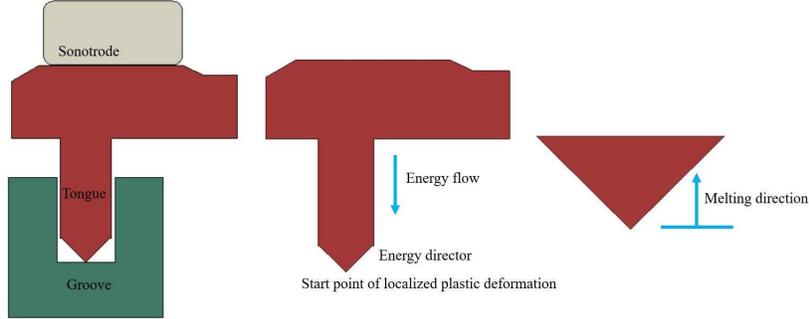}
\caption{
Triangle tongue and groove energy director design from Herrmann Ultrasonics, Inc.  Energy flows from tongue to energy director. Energy director melting direction will start from the tip of the triangle. 
}

\label{f:f3}
\end{figure}

In ultrasonic welding process, welding quality is controlled by many factors such as welding time, welding amplitude, ultrasonic frequency, design of booster and horn, and energy directors \citep{tao_influence_2019, yang_ultrasonic_2022}. Factors except energy directors are used to control how much ultrasonic wave energy flows to energy directors. This energy will finally go to energy directors, which will be melt and used to bond two adherends \citep{tao_influence_2019}.
 Energy directors can be designed as different shapes such as semicircular, triangular, rectangular \citep{bhudolia_advances_2020}. For example, Herrmann Ultrasonics, Inc. designed a triangular energy director,  shown in Fig.(\ref{f:f3}). Purpose of energy directors has two. The first one is to create localized plastic deformation and localized temperature rising. Consider Eq.(\ref{eq:localized}), when $E < R$, localized plastic deformation will be generated. Consider Eq.(\ref{eq:resistance1}), $R$ is a function of strain and stress. At the tip of a triangular energy director, strain and stress are the largest which means localized heat will start from the tip. The first melting point will start from the tip. This can also be confirmed by Eq.(\ref{eq:mech_resistance}). Mechanical resistance will increase if cross section of an energy director decreases. Therefore, the tip part of this energy director will have the largest mechanical resistance. Melting will be initiated at the tip.
 
The second purpose of an energy director is bond two adherends after melting. Ideally designed energy directors should be melted completely. Eq.(\ref{eq:ohm1}) can be used to calculate heat generated by ultrasonic wave. With particular design of energy directors, we can adjust ultrasonic amplitude and frequency, welding time to melt it completely. Since the mechanical resistance of an energy director is related to its length and cross section by Eq.(\ref{eq:mech_resistance}), the amount of heat generated will change with different shapes of energy director design \citep{chuah_effects_2000}. We can improve welding speed and obtain the same bonding strength by properly designed energy directors.

\section{Conclusion} 
\label{s:conclusion}
In this article, physics and mathematics of ultrasonic bonding process are illustrated. We are trying to understand the principle of ultrasonic welding process in mesoscale level or even in quantum level. We compared defect motions in mesoscale to electron motions in quantum level. Based on former work and history learning, we showed that they can share the same mathematical structures.
Since many mesoscale mechanisms of ultrasonic welding processes can hardly be captured by experimental measurements and finite element analysis, most of current work are based on physical intuition and literature study. With literature study,
energy of ultrasonic wave is mostly absorbed by defects in the materials. These defects will transfer their energy to lattices in crystalline structures or main backbone chains in amorphous structures. Large amplitude lattice vibration and main backbone chain vibration will lead to high temperature rising. When large amplitude vibration increases to a certain value, it is possible that covalent bonds of lattices or main backbone chains break apart. Materials will be melted. 
We also propose that an ultrasonic welding process is an inverse procedure to shear band formation. Both are due to localization of plastic deformation and temperature rising. Shear band will damage materials while ultrasonic bonding will repair materials.

\bibliographystyle{elsarticle-num}
\bibliography{xia_dft}

\begin{thebibliography}{10}
\expandafter\ifx\csname url\endcsname\relax
  \def\url#1{\texttt{#1}}\fi
\expandafter\ifx\csname urlprefix\endcsname\relax\def\urlprefix{URL }\fi
\expandafter\ifx\csname href\endcsname\relax
  \def\href#1#2{#2} \def\path#1{#1}\fi

\bibitem{barroeta_robles_repair_2022}
J.~Barroeta~Robles, M.~Dubé, P.~Hubert, A.~Yousefpour, Repair of thermoplastic
  composites: an overview, Advanced Manufacturing: Polymer \& Composites
  Science 8~(2) (2022) 68--96.

\bibitem{sun_highly_2012}
J.-Y. Sun, X.~Zhao, W.~R.~K. Illeperuma, O.~Chaudhuri, K.~H. Oh, D.~J. Mooney,
  J.~J. Vlassak, Z.~Suo, Highly stretchable and tough hydrogels, Nature
  489~(7414) (2012) 133--136.

\bibitem{gong_double-network_2003}
J.~Gong, Y.~Katsuyama, T.~Kurokawa, Y.~Osada, Double-{Network} {Hydrogels} with
  {Extremely} {High} {Mechanical} {Strength}, Advanced Materials 15~(14) (2003)
  1155--1158.

\bibitem{ahmadi_experimental_2022}
R.~Ahmadi, H.~Biglari, A.~Mostafapour, M.~Khoshravan, Experimental and
  numerical investigation of the traction-separation law of mode {II} fracture
  in single-edge ultrasonic welding in polypropylene composite reinforced by
  glass fibers, Journal of Adhesion Science and Technology 0~(0) (2022) 1--26.

\bibitem{slootman_quantifying_2020}
J.~Slootman, V.~Waltz, C.~J. Yeh, C.~Baumann, R.~Göstl, J.~Comtet, C.~Creton,
  Quantifying {Rate}- and {Temperature}-{Dependent} {Molecular} {Damage} in
  {Elastomer} {Fracture}, Physical Review X 10~(4) (2020) 041045.

\bibitem{zhou_high-throughput_2022}
Y.~Zhou, X.~Zhang, M.~Yang, Y.~Pan, Z.~Du, J.~Blanchet, Z.~Suo, T.~Lu,
  High-throughput experiments for rare-event rupture of materials, Matter 5~(2)
  (2022) 654--665.

\bibitem{unnikrishnan_review_2022}
T.~G. Unnikrishnan, P.~Kavan, A review study in ultrasonic-welding of similar
  and dissimilar thermoplastic polymers and its composites, Materials Today:
  Proceedings 56 (2022) 3294--3300.

\bibitem{li_ultrasonic_2022}
H.~Li, C.~Chen, R.~Yi, Y.~Li, J.~Wu, Ultrasonic welding of fiber-reinforced
  thermoplastic composites: a review, The International Journal of Advanced
  Manufacturing Technology 120~(1) (2022) 29--57.

\bibitem{bendikiene_study_2021}
R.~Bendikiene, L.~Kavaliauskiene, M.~Borkys, Study of ultrasonically welded
  thermoplastic dowel-wood board assembly, CIRP Journal of Manufacturing
  Science and Technology 35 (2021) 872--881.

\bibitem{jongbloed_differences_2020}
B.~Jongbloed, J.~Teuwen, R.~Benedictus, I.~F. Villegas, On differences and
  similarities between static and continuous ultrasonic welding of
  thermoplastic composites, Composites Part B: Engineering 203 (2020) 108466.

\bibitem{wagner_ultrasonic_2013}
G.~Wagner, F.~Balle, D.~Eifler, Ultrasonic {Welding} of {Aluminum} {Alloys} to
  {Fiber} {Reinforced} {Polymers}, Advanced Engineering Materials 15~(9) (2013)
  792--803.

\bibitem{yang_numerical_2022}
Y.~Yang, Z.~Liu, Y.~Wang, Y.~Li, Numerical {Study} of {Contact} {Behavior} and
  {Temperature} {Characterization} in {Ultrasonic} {Welding} of {CF}/{PA66},
  Polymers 14~(4) (2022) 683.

\bibitem{rubino_ultrasonic_2020}
F.~Rubino, H.~Parmar, V.~Esperto, P.~Carlone, Ultrasonic welding of magnesium
  alloys: a review, Materials and Manufacturing Processes 35~(10) (2020)
  1051--1068.

\bibitem{guo_joining_2019}
H.~Guo, M.~B. Gingerich, L.~M. Headings, R.~Hahnlen, M.~J. Dapino, Joining of
  carbon fiber and aluminum using ultrasonic additive manufacturing ({UAM}),
  Composite Structures 208 (2019) 180--188.

\bibitem{koutras_thermal_2021}
N.~Koutras, R.~Benedictus, I.~F. Villegas, Thermal effects on the performance
  of ultrasonically welded {CF}/{PPS} joints and its correlation to the degree
  of crystallinity at the weldline, Composites Part C: Open Access 4 (2021)
  100093.

\bibitem{zhang_study_2010}
Z.~Zhang, X.~Wang, Y.~Luo, Z.~Zhang, L.~Wang, Study on {Heating} {Process} of
  {Ultrasonic} {Welding} for {Thermoplastics}, Journal of Thermoplastic
  Composite Materials 23~(5) (2010) 647--664.

\bibitem{bhudolia_advances_2020}
S.~K. Bhudolia, G.~Gohel, K.~F. Leong, A.~Islam, Advances in {Ultrasonic}
  {Welding} of {Thermoplastic} {Composites}: {A} {Review}, Materials 13~(6)
  (2020) 1284.

\bibitem{levy_modeling_2014}
A.~Levy, S.~Le~Corre, I.~Fernandez~Villegas, Modeling of the heating phenomena
  in ultrasonic welding of thermoplastic composites with flat energy directors,
  Journal of Materials Processing Technology 214~(7) (2014) 1361--1371.

\bibitem{koellhoffer_role_2011}
S.~Koellhoffer, J.~W. Gillespie, S.~G. Advani, T.~A. Bogetti, Role of friction
  on the thermal development in ultrasonically consolidated aluminum foils and
  composites, Journal of Materials Processing Technology 211~(11) (2011)
  1864--1877.

\bibitem{villegas_ultrasonic_2019}
I.~F. Villegas, Ultrasonic {Welding} of {Thermoplastic} {Composites}, Frontiers
  in Materials 6.

\bibitem{yang_note_2018}
L.~Yang, L.~Yang, Note on {Gent}'s hyperelastic model, Rubber Chemistry and
  Technology 91~(1) (2018) 296--301.

\bibitem{de_gennes_reptation_1971}
P.~G. de~Gennes, Reptation of a {Polymer} {Chain} in the {Presence} of {Fixed}
  {Obstacles}, The Journal of Chemical Physics 55~(2) (1971) 572--579.

\bibitem{yang_mathematical_2019}
L.~Yang, A {Mathematical} {Model} for {Amorphous} {Polymers} {Based} on
  {Concepts} of {Reptation} {Theory}, Polymer Engineering \& Science 59~(11)
  (2019) 2335--2346.

\bibitem{gohel_effect_2022}
G.~Gohel, C.~Z. Soh, K.~F. Leong, P.~Gerard, S.~K. Bhudolia, Effect of {PMMA}
  {Coupling} {Layer} in {Enhancing} the {Ultrasonic} {Weld} {Strength} of
  {Novel} {Room} {Temperature} {Curable} {Acrylic} {Thermoplastic} to {Epoxy}
  {Based} {Composites}, Polymers 14~(9) (2022) 1862.

\bibitem{charlier_ultrasonic_2022}
Q.~Charlier, J.~Viguié, B.~Harthong, D.~Imbault, R.~Peyroux, P.~Martinez,
  M.~Caron, D.~Guérin, Ultrasonic welding of poly(vinyl alcohol) coated-papers
  hydrophobized by chromatogeny grafting, Cellulose 29~(18) (2022) 9939--9951.

\bibitem{b_g_brito_effects_2022}
C.~B.~G.~Brito, J.~Teuwen, C.~A. Dransfeld, I.~F.~Villegas, The effects of
  misaligned adherends on static ultrasonic welding of thermoplastic
  composites, Composites Part A: Applied Science and Manufacturing 155 (2022)
  106810.

\bibitem{frederick_disassembly_2021}
H.~Frederick, W.~Li, G.~Palardy, Disassembly {Study} of {Ultrasonically}
  {Welded} {Thermoplastic} {Composite} {Joints} via {Resistance} {Heating},
  Materials 14~(10) (2021) 2521.

\bibitem{siddiq_thermomechanical_2008}
A.~Siddiq, E.~Ghassemieh, Thermomechanical analyses of ultrasonic welding
  process using thermal and acoustic softening effects, Mechanics of Materials
  40~(12) (2008) 982--1000.

\bibitem{li_integrated_2020}
Y.~Li, T.~H. Lee, M.~Banu, S.~J. Hu, An integrated process-performance model of
  ultrasonic composite welding based on finite element and artificial neural
  network, Journal of Manufacturing Processes 56 (2020) 1374--1380.

\bibitem{yang_numerical_2010}
L.~Yang, R.~L. Lowe, S.-T.~J. Yu, S.~E. Bechtel, Numerical {Solution} by the
  {CESE} {Method} of a {First}-{Order} {Hyperbolic} {Form} of the {Equations}
  of {Dynamic} {Nonlinear} {Elasticity}, ASME Journal of Vibrations and
  Acoustics 132~(5) (2010) 051003.

\bibitem{yu_numerical_2010}
S.-T.~J. Yu, L.~Yang, R.~L. Lowe, S.~E. Bechtel, Numerical simulation of linear
  and nonlinear waves in hypoelastic solids by the {CESE} method, Wave Motion
  47~(3) (2010) 168--182.

\bibitem{lin_fracture_2021}
S.~Lin, J.~Ni, D.~Zheng, X.~Zhao, Fracture and fatigue of ideal polymer
  networks, Extreme Mechanics Letters 48 (2021) 101399.

\bibitem{yang_revisit_2020}
L.~Yang, L.~Yang, Revisit initiation of localized plastic deformation: {Shear}
  band \& necking, Extreme Mechanics Letters 40 (2020) 100914.

\bibitem{yang_viscoelasticity_2021}
L.~Yang, L.~Yang, R.~L. Lowe, A viscoelasticity model for polymers: {Time},
  temperature, and hydrostatic pressure dependent {Young}'s modulus and
  {Poisson}'s ratio across transition temperatures and pressures, Mechanics of
  Materials 157 (2021) 103839.

\bibitem{palardy_study_2018}
G.~Palardy, H.~Shi, A.~Levy, S.~Le~Corre, I.~Fernandez~Villegas, A study on
  amplitude transmission in ultrasonic welding of thermoplastic composites,
  Composites Part A: Applied Science and Manufacturing 113 (2018) 339--349.

\bibitem{li_weld_2019}
Y.~Li, Z.~Liu, J.~Shen, T.~H. Lee, M.~Banu, S.~J. Hu, Weld {Quality}
  {Prediction} in {Ultrasonic} {Welding} of {Carbon} {Fiber} {Composite}
  {Based} on an {Ultrasonic} {Wave} {Transmission} {Model}, Journal of
  Manufacturing Science and Engineering 141~(8).

\bibitem{yang_modeling_2011}
L.~Yang, Modeling {Waves} in {Linear} and {Nonlinear} {Solids} by
  {First}-{Order} {Hyperbolic} {Differential} {Equations}, {PhD} {Thesis}, The
  Ohio State University (2011).

\bibitem{rogale_interdependence_2022}
D.~Rogale, S.~Fajt, S.~FirÅ¡t~Rogale, Å.~Knezić, Interdependence of
  {Technical} and {Technological} {Parameters} in {Polymer} {Ultrasonic}
  {Welding}, Machines 10~(10) (2022) 845.

\bibitem{yang_viscoelasticity_2013}
L.~Yang, Y.-Y. Chen, S.-T.~J. Yu, Viscoelasticity determined by measured wave
  absorption coefficient for modeling waves in soft tissues, Wave Motion 50~(2)
  (2013) 334--346.

\bibitem{yang_theoretical_2020}
L.~Yang, Theoretical and numerical analysis of anterior cruciate ligament
  injury and its prevention, Global Journal of Research In Engineering.

\bibitem{long_experiments_2022}
T.~Long, S.~Shende, C.-Y. Lin, K.~Vemaganti, Experiments and hyperelastic
  modeling of porcine meniscus show heterogeneity at high strains, Biomechanics
  and Modeling in Mechanobiology 21~(6) (2022) 1641--1658.

\bibitem{bakavos_mechanisms_2010}
D.~Bakavos, P.~B. Prangnell, Mechanisms of joint and microstructure formation
  in high power ultrasonic spot welding 6111 aluminium automotive sheet,
  Materials Science and Engineering: A 527~(23) (2010) 6320--6334.

\bibitem{tao_influence_2019}
W.~Tao, X.~Su, H.~Wang, Z.~Zhang, H.~Li, J.~Chen, Influence mechanism of
  welding time and energy director to the thermoplastic composite joints by
  ultrasonic welding, Journal of Manufacturing Processes 37 (2019) 196--202.

\bibitem{yang_ultrasonic_2022}
Y.~Yang, Y.~Li, Z.~Liu, Y.~Li, S.~Ao, Z.~Luo, Ultrasonic welding of short
  carbon fiber reinforced {PEEK} with spherical surface anvils, Composites Part
  B: Engineering 231 (2022) 109599.

\bibitem{chuah_effects_2000}
Y.~K. Chuah, L.-H. Chien, B.~C. Chang, S.-J. Liu, Effects of the shape of the
  energy director on far-field ultrasonic welding of thermoplastics, Polymer
  Engineering \& Science 40~(1) (2000) 157--167.

\end{thebibliography}

\end{document}